\documentclass[journal,comsoc]{IEEEtran}
\usepackage[T1]{fontenc}

\ifCLASSINFOpdf
  \usepackage[pdftex]{graphicx}
  \graphicspath{{../pdf/}{../jpeg/}}
  \DeclareGraphicsExtensions{.pdf,.jpeg,.png}
\else
  \usepackage[dvips]{graphicx}
  \graphicspath{{../eps/}}
  \DeclareGraphicsExtensions{.eps}
\fi

\usepackage{amsmath}

\usepackage{bm}

\newtheorem{theorem}{Theorem}

\newtheorem{proposition}[theorem]{Proposition}
\usepackage{algorithm}  
\usepackage{algorithmicx}  
\usepackage{algpseudocode}

\usepackage{url}

\begin{document}
\bstctlcite{BSTcontrol}

\title{Transmit Power Optimization of Simultaneous Transmission and Reflection RIS Assisted Full-Duplex Communications}

\author{Yiru Wang, Pengxin Guan, Hongkang Yu and Yuping Zhao
}

\markboth{Journal of \LaTeX\ Class Files,~Vol.~14, No.~8, August~2015}%
{Shell \MakeLowercase{\textit{et al.}}: Bare Demo of IEEEtran.cls for IEEE Communications Society Journals}

\maketitle

\begin{abstract}
This work demonstrates the effectiveness of simultaneous transmission and reflection reconfigurable intelligent surface (STAR-RIS) aided Full-Duplex (FD) communication system. The objective is to minimize the total transmit power subject to the given minimal data rate requirement. We decouple the original problem into power optimization and STAR-RIS passive beamforming subproblems and adopt the alternating optimization (AO) framework to solve them. Specifically, in each iteration, we derive the closed-form expression for the optimal power design, then use the successive convex approximation (SCA) method and semidefinite program (SDP) to solve the passive beamforming optimization subproblem. Simulation results verify the convergence and effectiveness of the proposed algorithm, and further reveal the performance gain compared with the Half-Duplex (HD) and conventional RIS.
\end{abstract}

\begin{IEEEkeywords}
Simultaneously transmitting and reflecting reconfigurable intelligent surface, full-duplex, beamforming design.
\end{IEEEkeywords}

%
\IEEEpeerreviewmaketitle

\section{Introduction}

\IEEEPARstart{R}{ecently}, reconfigurable intelligent surface (RIS) has gained in popularity because of its potential to improve the wireless network performance \cite{RIS1}. RIS is equipped with large number of passive reflecting elements, which can dynamically change the wireless channels by adjusting the phase shifts and/or amplitude \cite{RIS2}. 

Considering the conventional RIS can either reflect or transmit signals, to improve the adaptivity of different communication scenarios, a novel technique called simultaneous transmission and reflection reconfigurable intelligent surface (STAR-RIS) was proposed in \cite{STAR1},  \cite{STAR360}. By altering the electromagnetic properties of the STAR-RIS elements with a smart controller, it can split the incident signal into transmission (T) region and the reflection (R) region, achieving 360$^o$ coverage. The STAR-RIS assisted non-orthogonal multiple access (NOMA) networks was studied in \cite{ref1}. To maximize the energy efficiency problem for STAR-RIS assisted network, a deep reinforcement learning method was proposed in \cite{Ding}. The author in \cite{ref3} also investigated the weighted sum secrecy rate in a STAR-RIS aided network.

On the other hand, the full-duplex (FD) communication has been studied to boost the spectral efficiency of wireless systems \cite{intro1}, \cite{intro2}. FD technology enables signal transmission and reception over the same time-frequency dimension and thus doubles the spectral efficiency theoretically compared with half-duplex (HD). Nevertheless, communications in FD mode would suffer from strong self-interference (SI) signal. Fortunately, a number of SI cancellation methods can suppress the SI power to the noise floor, which promotes the FD-based applications \cite{FDSI}.

This work study the transmit power optimization problem of the STAR-RIS aided FD communication system, where an uplink (UL) user and a downlink (DL) user locate on the opposite side of the STAR-RIS. To the best of our knowledge, this is the first work that studies the combination of STAR-RIS and FD. The main contributions are summarized as follows:

\begin{itemize}
	\item We study the transmit power minimization problem, subject to the minimum data rate demands of the UL and the DL, together with the transmission and reflection (T\&R) coefficients constraint at the STAR-RIS.
	\item We decouple the non-convex problem into power optimization and passive beamforming subproblems and use the alternative optimization (AO) framework to solve them iteratively, where the closed-form optimal power design is derived in every iteration. 
    \item To make the initial solution feasible for subsequent iterative optimization, we proposed an orthogonal interference transmission method (OITM) to initialize the transmitting coefficients of the STAR-RIS.
    \item The performance of STAR-RIS assisted FD system is compared with STAR-RIS assisted HD system and conventional-RIS assisted FD system. Simulation results verify the convergence and effectiveness of the proposed algorithm. 
\end{itemize}

\begin{figure}
	\centering
	\includegraphics[width=2.4in]{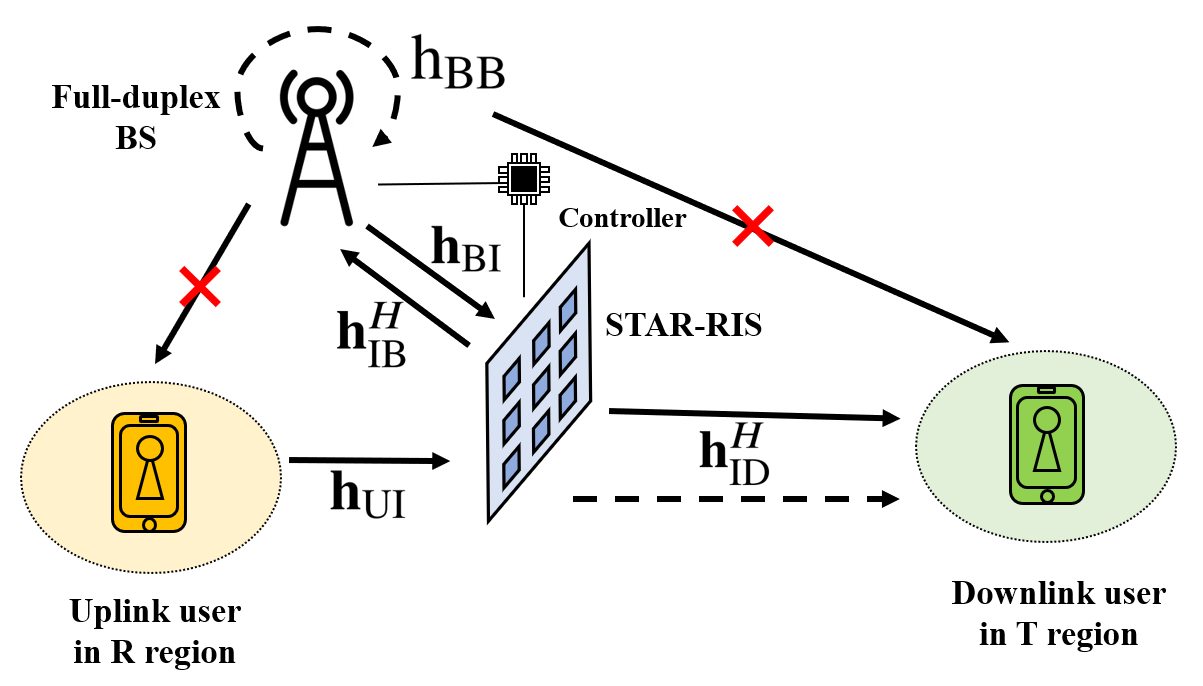}
	\caption{A STAR-RIS assisted FD communication system.}
\end{figure}

\emph{Notation}: $\left| x \right|$ and  ${\left\| {\bf{x}} \right\|_2}$ denote the absolution value of a scalar $x$ and Euclidean norm of a column vector ${\bf{x}} $. Tr(${\bf{X}}$), ${{\bf{X}}^T}$, ${{\bf{X}}^H}$, ${\rm{\uplambda }}({\bf{X}})$, and rank(${\bf{X}}$) denote the trace, transpose, conjugate transpose, maximum eigenvalue, and rank of the matrix ${\bf{X}}$, respectively. ${\left[ {\bf{x}} \right]_m}$ is the $(m)$-th element of vector ${\bf{x}}$. Diag(${\bf{x}}$) is a diagonal matrix with the entries of ${\bf{x}}$ on its main diagonal. ${\mathbb{C}^{m \times n}}$ denotes the space of ${m \times n}$ complex matrices.

\section{SYSTEM MODEL AND PROBLEM FORMULATION}

\subsection{STAR-RIS Description}
Assuming the total number of STAR-IRS transmission and reflection (T\&R) elements is $M$. Let ${s_m}$ denote the signal incident on the $m$-th element, where $m \in \left\{ {1,2,...,M} \right\}$. The transmitted signal and the reflected signal on the opposite side of STAR-RIS can be denoted as ${t_m} = \sqrt {\beta _m^t} {e^{j\phi _m^t}}{s_m}$ and ${r_m} = \sqrt {\beta _m^r} {e^{j\phi _m^r}}{s_m}$, where $\sqrt {\beta _m^t}  \in \left[ {0,1} \right]$ and $\sqrt {\beta _m^r}  \in \left[ {0,1} \right]$ denote the amplitude responses of the $m$-th element’s T\&R coefficients. To obey the law of energy conservation \cite{ref1}, we restrict that $\beta _m^t + \beta _m^r = 1$. $\phi _m^t \in \left[ {0,2\pi } \right)$ and $\phi _m^r \in \left[ {0,2\pi } \right)$ are the phase response of the $m$-th element’s T\&R coefficients.  We assume an ideal STAR-RIS with adjustable surface electric and magnetic impedance is deployed in the system, thus $\phi _m^t$ and $\phi _m^r$ can be tuned independently \cite{ref1}.

The transmitting coefficient vectors of the whole STAR-RIS can be expressed as 
\begin{equation}
{{\bf{q}}_t} = {(\sqrt {\beta _1^t} {e^{j\phi _1^t}},...,\sqrt {\beta _M^t} {e^{j\phi _M^t}})^T},
\end{equation}
and the reflecting coefficient vectors can be expressed as
\begin{equation}
{{\bf{q}}_r} = {(\sqrt {\beta _1^r} {e^{j\phi _1^r}},...,\sqrt {\beta _M^r} {e^{j\phi _M^r}})^T}.
\end{equation}

\subsection{System Model Description}
Consider a FD system consisting of a BS, a STAR-RIS, an UL user and a DL user, as depicted in Fig. 1. We assume that there is no direct link between the users and the BS because of deep fading or heavy shadowing. The BS is equipped with a single transmit antenna and a receive antenna, which operates in the FD mode. The UL and DL users are both equipped with a single antenna, and operate in the HD mode. We assume the STAR-RIS adopt the energy splitting (ES) protocol \cite{STAR1}, where all elements can simultaneously transmit and reflect signals. 

Denote ${{\bf{h}}_\text{UI}} \in {{\mathbb C}^{M \times 1}}$, ${\bf{h}}_\text{IB}^H \in {{\mathbb C}^{1 \times M}}$, ${{\bf{h}}_\text{BI}} \in {{\mathbb C}^{M \times 1}}$, ${\bf{h}}_\text{ID}^H \in {{\mathbb C}^{1 \times M}}$, ${{\text{h}}_\text{BB}} \in {\mathbb C}$ as the channel between the UL user and the STAR-RIS, between the STAR-RIS and the BS, between the BS and the STAR-RIS, between the STAR-RIS and the DL user and SI channel of the BS respectively. In addition, we assume perfect channel state information (CSI) can be obtained, thus the optimized system performance can serve as a lower bound. According to \cite{neg1}, \cite{neg2}, the reflected SI from the STAR-RIS can be reasonably neglected or cancelled. 

Let us denote $\sqrt {{p_U}} {x_U}$ as the signal transmitted from the UL user to the BS, where ${p_U}$ is the transmit power. Meanwhile, the BS transmits  $\sqrt {{p_D}} {x_D}$ to the DL user with given power ${p_D}$. Thus, the signal received at the BS can be expressed as
\begin{equation}
{y_U} = {\bf{h}}_\text{IB}^H{{\bm{\Phi }}_r}{{\bf{h}}_\text{UI}}\sqrt {{p_U}} {x_U} + \underbrace {{\text{h}_\text{BB}}\sqrt {{p_U}} {x_D}}_{\text{SI at the BS}} + {n_U},
\end{equation}
where ${{\bf{\Phi }}_r} = \text{Diag}({{\bf{q}}_r})$, and ${n_U} \sim CN(0,\sigma _U^2)$ is the additive white Gaussian noise (AWGN) at the BS. The received signal at the DL user can be expressed as
\begin{equation}\label{yd}
{y_D} = {\bf{h}}_\text{ID}^H{{\bm{\Phi }}_t}{{\bf{h}}_\text{BI}}\sqrt {{p_D}} {x_D} + \underbrace {{\bf{h}}_\text{ID}^H{{\bm{\Phi }}_t}{{\bf{h}}_\text{UI}}\sqrt {{p_U}} {x_U}}_{\text{interference from the UL user}} + {n_D},
\end{equation}
where ${{\bf{\Phi }}_t} = \text{Diag}({{\bf{q}}_t})$, and ${n_D} \sim CN(0,\sigma _U^2)$ is the AWGN at the DL user.

Therefore, the achievable data rate in bits second per Hertz (bps/Hz) of the UL and the DL can be formulated as 
\begin{equation}\label{ru}
{R_U} = {\log _2}\left( {1 + \frac{{{p_U}{{\left| {{\bf{h}}_\text{IB}^H{{\bm{\Phi }}_r}{{\bf{h}}_\text{UI}}} \right|}^2}}}{{{p_D}{{\left| {{\text{h}_\text{BB}}} \right|}^2} + \sigma _U^2}}} \right),
\end{equation}
and
\begin{equation}\label{rd}
{R_D} = {\log _2}\left( {1 + \frac{{{p_D}{{\left| {{\bf{h}}_\text{ID}^H{{\bm{\Phi }}_t}{{\bf{h}}_\text{BI}}} \right|}^2}}}{{{p_U}{{\left| {{\bf{h}}_\text{ID}^H{{\bm{\Phi }}_t}{{\bf{h}}_\text{UI}}} \right|}^2} + \sigma _D^2}}} \right).
\end{equation}

\subsection{Problem Formulation}
In this letter, we target to minimize the total transmit power by jointly designing the transmit power of the UL and the DL, together with optimizing the passive beamforming at the STAR-RIS. Besides, our design is within the constraints of the minimum data rate demands and STAR-IRS T\&R elements feasible set.

Mathematically, the optimization problem is derived as
\begin{subequations}\label{opt1}
	\begin{alignat}{2} 
		\mathcal{P}1:\quad& {\mathop {\min}\limits_{{{p_{U}}},{{p_{D}}}, {{\bf{q}}_k}}} &\ &{{p_{U}}} + {{p_{D}}}\label{opt1A} \\
		& \quad{\textrm {s.t.}}
		&&{R_U} \geqslant R_U^{th},{\text{ }}{R_D} \geqslant R_D^{th},\label{opt1B}\\
		&&&{{\bf{q}}_l} \in \mathcal{F},\forall l \in \left\{ {t,r} \right\}, \label{opt1C}
	\end{alignat}
\end{subequations}	
where $R_U^{th}$ and $R_D^{th}$ denote the minimum data rate of the UL and the DL. $\mathcal{F}$ denotes the feasible set for the STAR-RIS T\&R coefficient, in which $\sqrt {\beta _m^t}  \in \left[ {0,1} \right]$, $\sqrt {\beta _m^r}  \in \left[ {0,1} \right]$, $\beta _m^t + \beta _m^r = 1$, $\phi _m^t \in \left[ {0,2\pi } \right)$ and $\phi _m^r \in \left[ {0,2\pi } \right)$.

The optimization problem $\mathcal{P}1$ is non-trivial considering the non-convex data rate requirements. In addition, the power optimization and STAR-RIS's passive beamforming vectors are highly-coupled, which also makes the optimization intractable.

\section{TRANSMIT POWER MINIMIZATION ALGORITHM DESIGN}
In this section, we decouple the problem $\mathcal{P}1$ into power optimization and STAR-RIS passive beamforming design subproblems and adopt the AO method to solve them iteratively.

\subsection{Power Optimization With Given ${{\bf{q}}_t}$ and ${{\bf{q}}_r}$}

We define ${{\bf{h}}_1} = \text{Diag}({\bf{h}}_\text{IB}^H){{\bf{h}}_\text{UI}}$, ${{\bf{h}}_2} = \text{Diag}({\bf{h}}_\text{ID}^H){{\bf{h}}_\text{BI}}$, ${{\bf{h}}_3} = \text{Diag}({\bf{h}}_\text{ID}^H){{\bf{h}}_\text{UI}}$ and ${{\bf{H}}_1} = {{\bf{h}}_1}*{\bf{h}}_1^H$, ${{\bf{H}}_2} = {{\bf{h}}_2}*{\bf{h}}_2^H$, ${{\bf{H}}_3} = {{\bf{h}}_3}*{\bf{h}}_3^H$. To facilitate formulation, we define  ${{\bf{Q}}_t} = {{\bf{q}}_t}{\bf{q}}_t^H$ and ${{\bf{Q}}_r} = {{\bf{q}}_r}{\bf{q}}_r^H$.

Therefore, with given ${{\bf{q}}_t}$ and ${{\bf{q}}_r}$, the original problem $\mathcal{P}1$ can be transformed into power optimization subproblem as follows
\begin{subequations}\label{opt3}
	\begin{alignat}{2} 
	\mathcal{P}2:\quad& {\mathop {\min}\limits_{{{p_{U}}},{{p_{D}}}}} &\ &{{p_{U}}} + {{p_{D}}}\label{opt2A} \\
	& \quad{\textrm {s.t.}}
	&&{p_U}\text{Tr}\left( {{{\bf{Q}}_r}{{\bf{H}}_1}} \right) \geqslant R_U^{TH}\left( {{p_D}{{\left| {{\text{h}_\text{BB}}} \right|}^2} + \sigma _U^2} \right),\label{opt2B}\\
	&&&{p_D}\text{Tr}\left( {{{\bf{Q}}_t}{{\bf{H}}_2}} \right) \geqslant R_D^{TH}\left( {{p_U}\text{Tr}\left( {{{\bf{Q}}_t}{{\bf{H}}_3}} \right) + \sigma _D^2} \right),\label{opt2C}
	\end{alignat}
\end{subequations}
where $R_U^{TH} = {2^{R_U^{th}}} - 1$ and $R_D^{TH} = {2^{R_D^{th}}} - 1$.

\begin{proposition}
	Minimal transmit power can be obtained if and only if the UL and DL transmit signals at the minimal requiring rate. 
\end{proposition}
\begin{IEEEproof}
Suppose the UL and DL transmit signals at the minimal requiring rate. Thus, according to (\ref{opt2B}) and (\ref{opt2C}), the DL transmit power can be expressed as	
\begin{equation}\label{pD}
{\bar p_D} = \frac{{R_D^{TH}\left[ {R_U^{TH}\sigma _U^2\text{Tr}\left( {{{\bf{Q}}_t}{{\bf{H}}_3}} \right) + \text{Tr}\left( {{{\bf{Q}}_r}{{\bf{H}}_1}} \right)\sigma _D^2} \right]}}{{\text{Tr}\left( {{{\bf{Q}}_t}{{\bf{H}}_2}} \right)\text{Tr}\left( {{{\bf{Q}}_r}{{\bf{H}}_1}} \right) - R_D^{TH}R_U^{TH}{{\left| {{\text{h}_\text{BB}}} \right|}^2}\text{Tr}\left( {{{\bf{Q}}_t}{{\bf{H}}_3}} \right)}},
\end{equation}
and ${\bar p_U}$ can further be expressed as
\begin{equation}\label{pU}
{\bar p_u} = \frac{{R_U^{TH}\left[ {{{\bar p}_D}{{\left| {{\text{h}_\text{BB}}} \right|}^2} + \sigma _U^2} \right]}}{{\text{Tr}\left( {{{\bf{Q}}_r}{{\bf{H}}_1}} \right)}}.
\end{equation}

We then assume that the UL user transmits signals to the BS at higher data rate ${\tilde R_D}$. By substituting $R_D^{TH}$ with ${\tilde R_D}$ in (\ref{pD}), the obtained power ${\tilde p_D}$ is larger than ${\bar p_D}$ considering the denominator is decreased while the numerator is increased. Meanwhile, the ${\tilde p_U}$ is also increased. Therefore, the total transmit power is increased. Similar conclusion can also be made if UL data rate is higher than $R_U^{TH}$.
\end{IEEEproof}

\subsection{Passive Beamforming Optimization With Given ${p_D}$ and ${p_U}$}
With given power design  ${p_D}$ and ${p_U}$, the STAR-RIS passive beamforming design subproblem is a feasibility check problem, which can be written as follows
\begin{subequations}\label{opt5}
	\begin{alignat}{2} 
	\mathcal{P}3:\;& {\rm{find}} &\ & {{{\bf{Q}}_t},{{\bf{Q}}_r}} \label{opt3A}\\
	& \;\;\;{\textrm {s.t.}}
	&&\text{Rank}\left( {{{\bf{Q}}_t}} \right) = 1,\text{Rank}\left( {{{\bf{Q}}_r}} \right) = 1,\label{opt3B}\\
	&&&(\ref{opt1C}), (\ref{opt2B})-(\ref{opt2C}),\label{opt3C}
	\end{alignat}
\end{subequations}

Problem $\mathcal{P}3$ is non-convex because of the rank-one constraint. Conventionally, we can use the semidefinite relaxation (SDR) method to drop the rank-one constraint and adopt gaussian randomization approach to obtain ${\bf{q}}_t^*$ (${\bf{q}}_t^*$). However, after relaxing the rank-one constraint, the solutions are not guaranteed to meet the data rate constraints. Hence, we develop a more efficient algorithm to find an optimal rank-one solution based on SCA method.

Based on the Proposition 3 in \cite{ref4}, the rank-one constraint (\ref{opt3B}) is satisfied when ${\bf{Q}}_t^*$ (${\bf{Q}}_t^*$) has only one non-zero eigenvalue. Thus, the rank-one constraint can be transformed as
\begin{equation}
\text{Tr}\left( {{{\bf{Q}}_l}} \right) - \uplambda \left( {{{\bf{Q}}_l}} \right) = 0.
\end{equation}
where $l \in \{ t,r\} $ represents any of the transmit or reflect mode. We define $f\left( {{{\bf{Q}}_l}} \right) = \text{Tr}\left( {{{\bf{Q}}_l}} \right) - \uplambda \left( {{{\bf{Q}}_l}} \right)$. Therefore, we can rewrite problem $\mathcal{P}3$ as
\begin{subequations}\label{opt6}
	\begin{alignat}{2} 
	\mathcal{P}4:\quad& {\mathop {\min}\limits_{{{{\bf{Q}}_t},{{\bf{Q}}_r}}}} &\ &f\left( {{{\bf{Q}}_t}} \right)\text{ + }f\left( {{{\bf{Q}}_r}} \right) \label{opt6A} \\
	& \quad{\textrm {s.t.}}
	&&(\ref{opt1C}), (\ref{opt2B})-(\ref{opt2C}).\label{opt6B}
	\end{alignat}
\end{subequations}

In order to ensure the convergence of SCA algorithm, we further rewrite $f\left( {{{\bf{Q}}_l}} \right)$ as $f\left( {{{\bf{Q}}_l}} \right) = \text{Tr}\left( {{{\bf{Q}}_l}} \right) + \frac{\rho }{2}\left\| {{{\bf{Q}}_l}} \right\|_F^2 - \left( {\uplambda \left( {{{\bf{Q}}_l}} \right) + \frac{\rho }{2}\left\| {{{\bf{Q}}_l}} \right\|_F^2} \right)$. $\frac{\rho }{2}\left\| {{{\bf{Q}}_l}} \right\|_F^2$ is the quadratic term which makes the objective function $f\left( {{{\bf{Q}}_l}} \right)$ to be the difference of two $\beta$-strongly convex functions. We define $g\left( {{{\bf{Q}}_l}} \right) = \left( {\uplambda \left( {{{\bf{Q}}_l}} \right) + \frac{\rho }{2}\left\| {{{\bf{Q}}_l}} \right\|_F^2} \right)$ , thus its first-order approximation at the feasible point ${\bf{Q}}_l^{\left( {k - 1} \right)}$ can be given by
\begin{equation}
g\left( {{{\bf{Q}}_l}} \right) \geqslant g\left( {{\bf{Q}}_l^{\left( {k - 1} \right)}} \right) + \text{Tr}\left\{ {\operatorname{Re} \left[ {\partial _{{\bf{Q}}_l^{\left( {k - 1} \right)}}^Hg\left( {{{\bf{Q}}_l}} \right)\left( {{{\bf{Q}}_l} - {\bf{Q}}_l^{\left( {k - 1} \right)}} \right)} \right]} \right\}.
\end{equation}
where ${\bf{Q}}_l^{\left( {k - 1} \right)}$ is the solution at the  $k$-th iteration. Re($x$) represents the real part of the complex number. 

The subgradient of $g\left( {{{\bf{Q}}_l}} \right)$ an be expressed as ${\partial _{{\bf{Q}}_l^{\left( {k - 1} \right)}}}g\left( {{{\bf{Q}}_l}} \right) = \alpha \left( {{\bf{Q}}_l^{\left( {k - 1} \right)}} \right)\alpha {\left( {{\bf{Q}}_l^{\left( {k - 1} \right)}} \right)^H} + \rho {\bf{Q}}_l^{\left( {k - 1} \right)}$, in which the $\alpha \left( {{\text{Q}}_l^{\left( {k - 1} \right)}} \right)$ is the corresponding eigenvector of the largest eigenvalue.

Hence, by adopting SCA method, the solution ${{\bf{Q}}_t}$ and ${{\bf{Q}}_r}$ at the $k$-th iteration can be obtained by solving the following problem
\begin{subequations}\label{opt6}
	\begin{alignat}{2} 
	\mathcal{P}4^{\prime}:\quad& {\mathop {\min}\limits_{{{{\bf{Q}}_t},{{\bf{Q}}_r}}}} &\ 
	&
	\sum\limits_l {\text{Tr}\left( {{{\bf{Q}}_l}} \right) + \frac{\rho }{2}\left\| {{{\bf{Q}}_l}} \right\|_F^2 
		- \text{Tr}\left\{ {\operatorname{Re} \left[ {\partial _{{\bf{Q}}_l^{\left( {k - 1} \right)}}^Hg\left( {{{\bf{Q}}_l}} \right){{\bf{Q}}_l}} \right]} \right\}} \\
	& \quad{\textrm {s.t.}}
	&&(\ref{opt1C}), (\ref{opt2B})-(\ref{opt2C}).\label{opt6B}
	\end{alignat}
\end{subequations}

Problem $\mathcal{P}4^{\prime}$ is a standard SDP that can be solved by CVX. If the object function iteratively decreases and is eventually below a given threshold $\varepsilon $, then we consider our proposed method succeeds in finding an optimal rank-one solution. Finally, we can get the optimal ${\bf{q}}_t^*$ and ${\bf{q}}_r^*$ by eigenvalue decomposition.

\subsection{Orthogonal Interference Transmission Method (OITM)}

In general, the solutions of ${p_U}$ and ${p_D}$ need to be positive. However, if we adopt the common-used random phase initiation for STAR-RIS, it would probably lead to negative value. To sidestep this failure, we propose an OITM scheme for phase initiation. 

\begin{proposition}
One initialization of ${{\bf{q}}_t}$ can be expressed as
	
\begin{equation}\label{qt0}
{{\bf{q}}_{{t_0}}} = \frac{{\left( {{{\bf{I}}_M} - {{\bf{h}}_3}{\bf{h}}_3^H/\left\| {{{\bf{h}}_3}} \right\|_2^2} \right){\bf{x}}}}{{\mathop {\max }\limits_m {{\left[ {\left( {{{\bf{I}}_M} - {{\bf{h}}_3}{\bf{h}}_3^H/\left\| {{{\bf{h}}_3}} \right\|_2^2} \right){\bf{x}}} \right]}_m}}},
\end{equation}
where $m \in \left\{ {1,...,M} \right\}$ and ${\text{x}} \in {\mathbb{C}^{M \times 1}}$ is a randomly generated unit-modulus vector. ${\bf{I}}$ represents an identity matrix.

\end{proposition}
\begin{IEEEproof}
When $\text{Tr}\left( {{{\bf{Q}}_t}{{\bf{H}}_3}} \right) = 0$, we can eliminate the negative occurrence of ${P_D}$ and ${P_U}$ according to (\ref{pD}) and (\ref{pU}). Therefore, the phase shift of transmitting elements ${{\bf{q}}_t}$ should be orthogonal to the interference channel ${{{\bf{h}}_3}}$. The orthogonal compliment projector of ${{{\bf{h}}_3}}$ can be given as

\begin{equation}
{\bf{P}}_A^ \bot  = {{\bf{I}}_M} - {{\bf{h}}_3}{\bf{h}}_3^H/\left\| {{{\bf{h}}_3}} \right\|_2^2.
\end{equation}

Hence, by generating one unit-modulus vector ${\bf{x}} \in {\mathbb{C}^{M \times 1}}$ we can obtain the transmitting coefficient vector as 
\begin{equation}
{{\bf{q}}_{{t_0}}} = \frac{{\bf{y}}}{{\mathop {\max }\limits_m {{\bf{y}}_m}}} = \frac{{\left( {{{\bf{I}}_M} - {{\bf{h}}_3}{\bf{h}}_3^H/\left\| {{{\bf{h}}_3}} \right\|_2^2} \right){\bf{x}}}}{{\mathop {\max }\limits_m {{\left[ {\left( {{{\bf{I}}_M} - {{\bf{h}}_3}{\bf{h}}_3^H/\left\| {{{\bf{h}}_3}} \right\|_2^2} \right){\bf{x}}} \right]}_m}}}.
\end{equation}

Hence, the phase shift of ${{\bf{q}}_t}$ can be initialized randomly and its element amplitude can be expressed as 
\begin{equation}\label{qr0}
\left| {{{\left[ {{{\bf{q}}_{{r_0}}}} \right]}_m}} \right| = 1 - \left| {{{\left[ {{{\bf{q}}_{{t_0}}}} \right]}_m}} \right|,
\end{equation}
\end{IEEEproof}
\subsection{Overall Algorithm}
Our proposed algorithm is summarized in Algorithm 1. ${\varepsilon _1}$ and ${\varepsilon _2}$ are small thresholds, while $N$ and $K$ represent the maximum numbers of iterations.

\begin{proposition}
	It can be guaranteed that our algorithm can converge to a local optimal solution.
\end{proposition}

\begin{IEEEproof}
	Without loss of generality, we suppose that ${P}(p^{\left( n \right)},{{\bf{q}}^{\left( n \right)}})$ is the total transmit power with given power design ${p^{\left( n \right)}} = p_U^{\left( n \right)} + p_D^{\left( n \right)}$ and passive STAR-RIS beamforming vectors ${{\bf{q}}^{\left( n \right)}} = \left( {{\bf{q}}_t^{\left( n \right)},{\bf{q}}_r^{\left( n \right)}} \right)$ at the  $n$-th iteration. Thus, we have the following formulation
	
\begin{equation}
{P}(p^{\left( n \right)},{{\bf{q}}^{\left( n \right)}})\mathop  \geqslant \limits^{\left( a \right)} {P}(p^{\left( {n + 1} \right)},{{\bf{q}}^{\left( n \right)}})\mathop = \limits^{\left( b \right)} {P}(p^{\left( {n + 1} \right)},{{\bf{q}}^{\left( {n + 1} \right)}}).
\end{equation}	
where $\left( a \right)$ holds because the power design ${p^{\left( {n + 1} \right)}} = p_U^{\left( {n + 1} \right)} + p_D^{\left( {n + 1} \right)}$ is optimal and updated at $(n+1)$-th iteration with the passive beamforming ${{\bf{q}}^{\left( n \right)}}$. However, the updated passive beamforming ${{\bf{q}}^{\left( {n + 1} \right)}}$ does not change the power design of ${p^{\left( {n + 1} \right)}}$ at $(n+1)$-th iteration, thus $\left( b \right)$ also holds.
\end{IEEEproof}

The complexity of the proposed algorithm is dominated by solving SDP $\mathcal{P}4^{\prime}$, for which the complexity is ${\rm O}\left( {{M^4}} \right)$. Hence, the overall complexity is ${\rm O}\left( {{I_1}{I_2}{M^4}} \right)$, where ${I_1}$ and ${I_2}$ are the iteration numbers of AO algorithm and SCA algorithm, respectively.

\begin{algorithm}[h]
	\caption{Alternating Optimization Algorithm for $\mathcal{P}1$}
\begin{algorithmic}[1]
	\State {\textbf{Initialization:}} set $n = 1$, initialize ${{\bf{q}}_t}$ based on (\ref{qt0}) and ${{\bf{q}}_r}$ based on (\ref{qr0}). 
	\Repeat
	\State Solve problem $\mathcal{P}2$ to get $p_U^{\left( n \right)}$ and $p_D^{\left( n \right)}$ with given ${{\bf{q}}_t^{\left( {n - 1} \right)}}$ and ${{\bf{q}}_r^{\left( {n - 1} \right)}}$;
	\State Set $k=0$;
	\While {$\rm{Tr}\left( {{\bf{Q}}_t^k} \right) - \uplambda \left( {{\bf{Q}}_t^k} \right) + \rm{Tr}\left( {{\bf{Q}}_r^k} \right) - \uplambda \left( {{\bf{Q}}_r^k} \right) \ge {\varepsilon _1}$ and $k \le K$}
	\State Solve $\mathcal{P}4^{\prime}$ to update ${{\bf{Q}}_t^{\left( {k + 1} \right)}}$ and ${{\bf{Q}}_r^{\left( {k + 1} \right)}}$ with given $p_U^{\left( n \right)}$ and $p_D^{\left( n \right)}$;
	\State Update $k=k+1$;
	\EndWhile
	\State Update ${{\bf{q}}_t^{\left( n \right)}}$ and ${{\bf{q}}_t^{\left( n \right)}}$ by eigenvalue decomposition;
	\State Calculate ${p^{\left( n \right)}} = p_U^{\left( n \right)} + p_D^{\left( n \right)}$;
	\State Update $n=n+1$;
	\Until ${{({P^{(n - 1)}} - {P^{(n)}})} \mathord{\left/
			{\vphantom {{({P^{(n - 1)}} - {P^{(n)}})} {{P^{(n - 1)}}}}} \right.
			\kern-\nulldelimiterspace} {{P^{(n - 1)}}}} \le {\varepsilon _2}$ or $n \ge N$.
	\State \textbf{Output:} power design and passive beamforming vectors.
\end{algorithmic}
\end{algorithm}
\section{SIMULATION RESULTS}
This section provides simulation results to verify the performance of our proposed algorithm. We assume that the locations of BS, STAR-RIS, UL user and DL user are (5m, 45m), (0m,50m), (0m, 35m) and (0m, 100m). We set $\sigma _U^2 = \sigma _D^2 =  - 80$dBm. The large-scale fading is modelled by $PL\left( d \right) = P{L_0}{\left( {d/{d_0}} \right)^{ - \varpi }}$,  where $P{L_0} =  - 30$dB is the path loss at the reference distance ${d_0} = 1$m, $d$ is the distance, and $\varpi $ is the path-loss exponent which is set to 2.2 \cite{STAR1}. We adopt the Rician model for all channels, where the Rician factor is 5dB in SI channel \cite{neg1} and 3dB for others \cite{STAR1}. The path loss of SI channel is much lower due to SI cancellation. We assume ${R_U}$ is 1 bps/Hz for all scenarios.

The proposed optimization scheme for the STAR-RIS aided FD system, namely STAR-RIS-FD, is compared to the following benchmark schemes:

\begin{itemize}
	\item STAR-RIS aided HD system (STAR-RIS-HD): We assume that the UP and DL communication is assigned with equal time slot. Then the T\&R coefficients in different slots are optimized separately.
	\item Conventional-RIS aided FD system (CON-RIS-FD): We assume a transmitting-only RIS and a reflecting-only RIS are adjacent to each other and deployed at the same coordinate as the STAR-RIS \cite{STAR1}, the total elements number is $M$. The UL communication is facilitated by the reflecting-only RIS, while the DL communication depends on the transmitting-only RIS. The system operates in FD mode.
\end{itemize}
\begin{figure}
	\centering
	\includegraphics[width=2.4in]{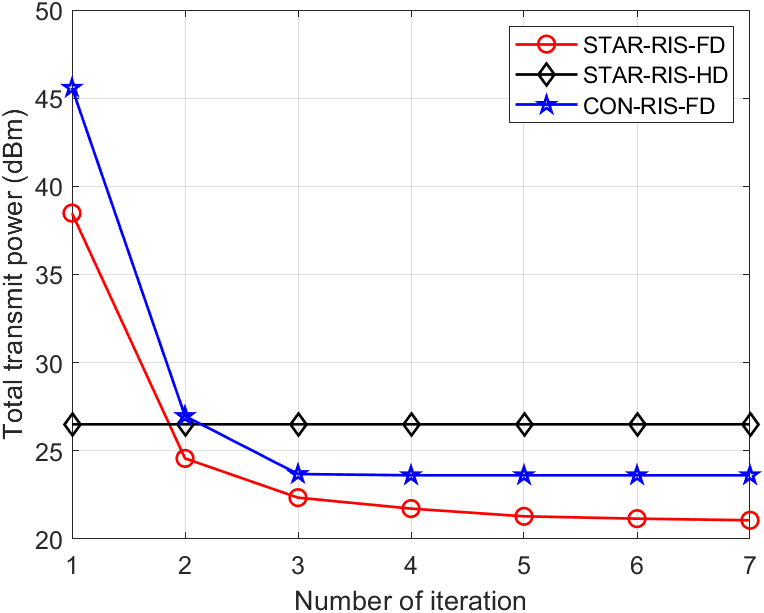}
	\caption{Performance of the proposed method versus the number of iterations.}
\end{figure}

Fig. 2 investigates the convergence of the proposed algorithm, where $M$ is set to 40, ${R_D}$ is set to 4 bps/Hz and the path loss of SI channel at the BS is set to -100dB. Considering the optimal power design of the STAR-RIS-HD scheme can be expressed in  closed-form expression, it converges at the first iteration. It can be seen that as the number of iterations continues to increase, the transmit power of STAR-RIS-FD and CON-RIS-FD schemes decreases to convergence. 

\begin{figure}
	\centering
	\includegraphics[width=2.4in]{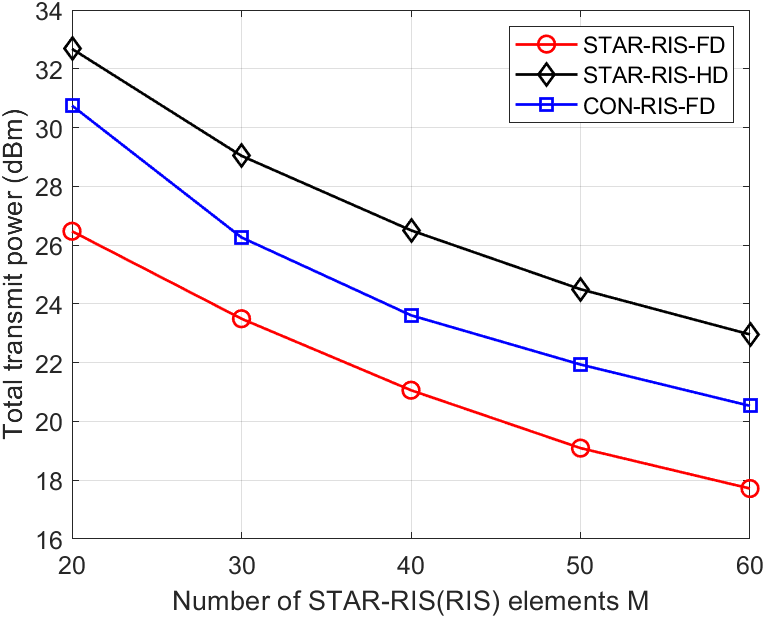}
	\caption{Total transmit power versus the number of STAR-RIS (CON-RIS) elements.}
\end{figure}

Fig. 3 shows the transmit power versus the number of STAR-RIS(CON-RIS) elements, where ${R_D}$ is set to 4 bps/Hz and the path loss of SI channel at the BS is set to -100dB. The minimal transmit power of all three schemes decreases with the increase of STAR-RIS (CON-RIS) elements. Compared to CON-RIS with fixed numbers of T\&R elements, STAR-RIS can exploit all degrees-of-freedom (DoFs) to enhance the desired signal strength and mitigate interference. Thus, the STAR-RIS outperforms the CON-RIS. Besides, the achievable rate is penalized due to the half communication time in the HD mode. Therefore, under the same data rate requirement in this scenario, the HD system consumes more transmit power compared with FD.

\begin{figure}
	\centering
	\includegraphics[width=2.4in]{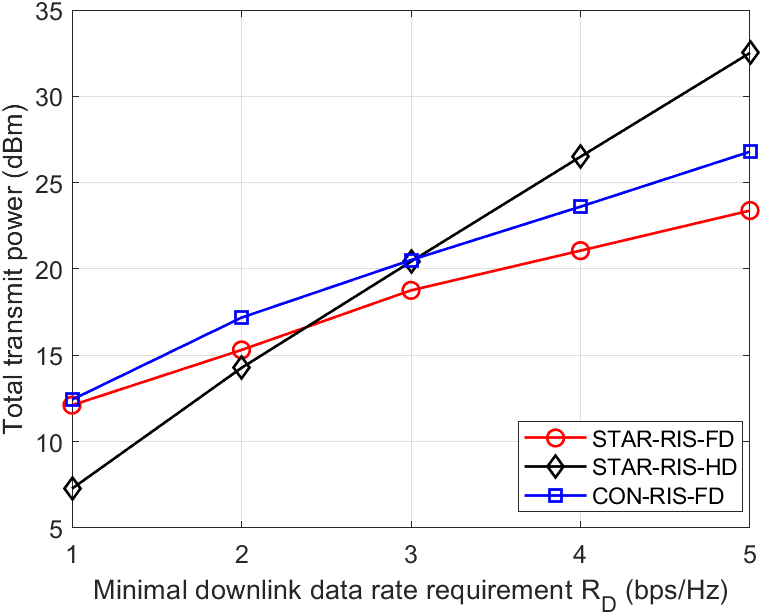}
	\caption{Total transmit power versus DL data rate requirement.}
\end{figure}

Fig. 4 shows the transmit power versus different DL data rate requirement, where $M$ is set to 40 and the path loss of SI is set to -100dB. When the DL data rate demand is lower than 3 bps/Hz, the interference from the UL has greater influence on the DL user than DL transmission from the BS. Therefore, the STAR-RIS uses its full capacity to restrain the DL interference. Interference causes much more deterioration to the FD mode compared with the penalty because of half communication time to the HD mode. However, as the DL data rate demand augments, the signal strength from the BS is much stronger than the interference. Thus, aiding the DL transmission is the STAR-RIS’s working priority, which is the same as DL transmission in HD mode. Besides, the half-time penalty to the HD mode increases rapidly with rate demand. Hence, the STAR-RIS in FD mode, even the CON-RIS working at FD mode, requires lower transmit power with higher DL rate demand.

\begin{figure}
	\centering
	\includegraphics[width=2.4in]{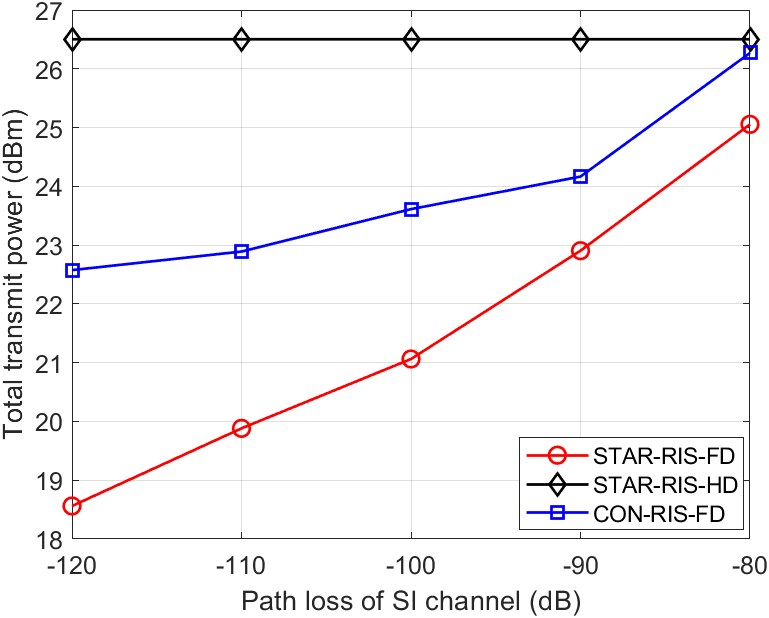}
	\caption{Total transmit power versus the path loss of SI channel.}
\end{figure}

Fig. 5 shows the transmit power versus the pass loss of SI at the BS, where $M$ is set to 40, ${R_D}$ is set to 4 bps/Hz. The SI intensity has no impact on the STAR-RIS system operating in HD. However, for STAR-RIS and CON-RIS working in the FD mode, the transmit power decreases with SI intensity. 

\section{CONCLUSION}
In this letter, we studied the STAR-RIS aided FD system. We formulated the transmit power minimization problem, subject to the minimum data rate demand of the UL and DL. We divided the optimization problem into power design and passive beamforming design subproblems, and adopted the AO framework to solve them. The closed-form optimal power design scheme was derived in every iteration. An OITM scheme was also proposed to initialize the transmitting coefficients of the STAR-RIS. Simulation results demonstrated the performance of the proposed algorithm and further revealed STAR-RIS assisted FD scheme outperforms HD mode in higher data rate and lower SI scenarios.

\ifCLASSOPTIONcaptionsoff
  \newpage
\fi

\bibliographystyle{IEEEtran}
\bibliography{IEEEabrv,myre}

\end{document}